\documentclass[12pt]{elsart}
\usepackage{epsfig,graphics}

\def\lsim{\mathrel{\rlap{\lower4pt\hbox{\hskip1pt$\sim$}}
    \raise1pt\hbox{$<$}}}         
\def\gsim{\mathrel{\rlap{\lower4pt\hbox{\hskip1pt$\sim$}}
    \raise1pt\hbox{$>$}}}         

\def\be{\begin{equation}}
\def\ee{\end{equation}}
\def\bq{\begin{eqnarray}}
\def\eq{\end{eqnarray}}
\def\bm{\boldmath}

\begin{document}
\pagestyle{empty}

\vspace{2.0cm}
 
\begin{center}
 
{\large \bf  FLAVOR ASYMMETRY OF THE \medskip \\ 
POLARIZED LIGHT SEA: MODELS VS. DATA \\}

\vspace{1.0cm}

{\large V.~Barone$^{a}$, T.~Calarco$^{b}$, 
A.~Drago$^{c}$ and M.C.~Simani$^{d}$\\}

\vspace{1.0cm} 

{\it $^{a}$Di.S.T.A., 
Universit{\`a} del Piemonte Orientale ``A.~Avogadro'', \\
and INFN, Gruppo Coll. di Alessandria,
15100 Alessandria, Italy 
\medskip \\
$^{b}$ ECT*, Villa Tambosi, I-38050 Villazzano (Trento), Italy
\medskip \\
$^{c}$Dipartimento di Fisica, Universit{\`a} di Ferrara, \\
and INFN, Sezione di Ferrara, 44100 Ferrara, Italy 
\medskip\\ 
$^{d}$ Lawrence Livermore National Laboratory, Livermore, CA 94550, USA
\medskip \\}

\vspace{1.0cm}

{\large \bf Abstract \bigskip\\ }

\end{center}

\noindent 
The flavor asymmetry 
of the polarized light sea, 
$\Delta \bar u - \Delta \bar d$,  
discriminates between different model calculations of 
helicity densities. 
We show that the chiral chromodielectric model,
differently from models based on a $1/N_c$ expansion, predicts 
a small value for this asymmetry, what seems in agreement  
with preliminary HERMES data.

\vfill
 
\pagebreak

\baselineskip 20 pt
\pagestyle{plain}

\noindent
1. Following the discovery that the quark  
contribution to the spin of the nucleon is 
surprisingly small \cite{ashman}, a considerable  
experimental effort was made 
to elucidate the details of the helicity densities 
of valence, sea and glue (for reviews 
on longitudinal spin physics, see \cite{reviews}). 
On the other theoretical side, many models have been
studied and most of them
reproduce the gross features 
of the spin content of the nucleon, namely the 
$g_1$ structure function and the singlet 
distribution $\Delta \Sigma 
= \sum_f (\Delta f + \Delta \bar f)$. 
In order to discriminate 
between the models, one has to look at  
the quark and antiquark helicity 
densities for each separate flavor.   
This has been a lacking piece of 
information until last year, when 
the HERMES 
Collaboration at DESY, measuring semi-inclusive deep 
inelastic scattering, succeeded in extracting  
the polarizations of $u, \bar u, d, \bar d, 
s + \bar s$ \cite{beckmann}. Thus, HERMES experiment opened for the first
time the possibility to test model results against data.   
To this purpose, non-singlet distributions are 
especially interesting because their evolution 
does not involve the polarized gluon density:
as a consequence, they can be predicted in a more reliable 
way, without any extra assumption on the 
constituent polarized glue.

In what follows, our attention will be directed    
to the isoscalar and isovector combinations of 
antiquark densities, 
\bq
& & \bar f_+(x) = \bar u (x) + \bar d (x)\,, 
\;\;\;\qquad\quad
\bar f_-(x) = \bar u (x) - \bar d (x)\,, 
\label{unpol} \\
& & \Delta \bar f_+(x) = \Delta \bar u (x) + \Delta \bar d (x)\,, 
\;\;\;\;\;
\Delta \bar f_-(x) = \Delta \bar u (x) - \Delta \bar d (x)\,. 
\label{pol}
\eq
The two classes of models  
most widely used for computing quark distributions, 
{\it i.e.} the chiral quark soliton 
model (CQSM, based on a $1/N_c$ expansion) 
and the bag-like confinement models -- including the 
chiral chromodielectric model (CDM) --, 
make very different predictions 
for the relative weight of $\Delta \bar f_-$ and 
$\Delta \bar f_+$, and of $\bar f_+$ and 
$\bar f_-$. In particular, in the $1/N_c$ expansion,  
$\Delta \bar f_-$ is a leading quantity compared to 
$\Delta \bar f_+$ and 
$\bar f_-$, hence it is expected to be large (in absolute 
value) and to satisfy the inequalities
(which may be, as a matter of fact,  
strong inequalities)
\bq
& & \vert \Delta \bar u - \Delta \bar d \vert 
> \vert \bar u - \bar d \vert\,, 
\label{ineq1} \\
& & \vert \Delta \bar u - \Delta \bar d \vert 
> \vert \Delta \bar u + \Delta \bar d \vert\,.
\label{ineq2}
\eq
On the contrary, we shall show that the chromodielectric 
model predicts a small value for the polarized 
flavor asymmetry $\Delta f_-$, and reversed  
signs for the inequalities (\ref{ineq1}, \ref{ineq2}).     
The HERMES preliminary data, although 
affected by relatively large uncertainties, seem 
indeed to favor 
a small $\Delta f_-$.

\vspace{0.5cm}

\noindent
2. Let us start from the field-theoretical 
expressions of the 
quark and antiquark helicity distributions, {\it i.e.}  
\bq
\Delta  f(x) &=& \int \frac{{\rm d}\xi^-}{4 \pi} 
e^{i x p^+\xi^-}
\langle N |
\overline{\psi}(0) \gamma^+ \gamma_5  
\psi (0, \xi^{-}, 0_{\perp})| N \rangle 
\,, 
\label{defhel1} \\
\Delta  \bar f(x) &=& \int \frac{{\rm d}\xi^-}{4 \pi} 
e^{i x p^+\xi^-}
\langle N |{\rm Tr} \, \gamma^+ \gamma_5
\psi(0)  
\overline{\psi} (0, \xi^{-}, 0_{\perp})| N \rangle \,.  
\label{defhel2} 
\eq
Quark models provide the matrix elements in the nucleon state, 
which cannot be calculated in perturbative QCD.  

In a (projected) mean-field approximation,
eqs.~(\ref{defhel1}, \ref{defhel2}) can be rewritten
in terms of single--particle quark or antiquark matrix elements. 
For the quark distribution one 
has \cite{thomas,noi,noi2} (the expression 
for antiquarks is similar)
\bq
\Delta &f(x)& = \frac{1}{\sqrt{2}}
\sum_\alpha\sum_m P(f,\alpha,m) 
\nonumber \\
 & \times &\, \int  
\frac{{\rm d}^3\mbox{\bm $p$}_\alpha}{(2\pi)^3 \, 2 p_\alpha^0}\,
A_\alpha(p_\alpha) \, \delta[(1-x)\, p^+ - p^+_\alpha]
\, \overline\varphi(p_\alpha,m)
\gamma^+ \gamma_5 \, \varphi(p_\alpha,m)
\, ,
\label{deltaf}
\eq
where $\varphi$ is the single-quark wave function, $m$
is the projection of the quark spin  along the direction of the nucleon's
spin, $P(f, \alpha, m)$ is the probability of extracting a quark of
flavor $f$ and spin $m$ leaving a state generically labeled by the 
quantum number $\alpha$.
The overlap function $A_\alpha(p_\alpha)$ 
contains the details of the intermediate states and 
of the projection used to obtain a
nucleon with definite linear momentum 
from a three--quark bag (see for instance \cite{thomas,noi}).
The intermediate states which contribute to 
(\ref{deltaf}) are $2q$ and $3q 1\bar q$ states
for the quark distribution, and $4q$ states for the antiquark distribution.

The model of the nucleon that we adopt 
is the chiral chromodielectric 
model (CDM) \cite{Pirner}. The Lagrangian of the CDM is 
\begin{eqnarray}
{\mathcal L} &=& i\bar \psi \gamma^{\mu}\partial_{\mu} \psi
       +{g\over \chi} \, \bar \psi\left(\sigma
+i\gamma_5 \mbox{\bm $\tau \cdot \pi$} \right) \psi        \nonumber
 \\
         &+&{1\over 2}{\left(\partial_\mu\chi\right)}^2
                       -{1\over 2} M^2\chi^2
                       +{1\over 2}{\left(\partial_\mu\sigma \right)}^2
             +{1\over 2}{\left(\partial_\mu \mbox{\bm $\pi$}\right)}^2
                       -U\left(\sigma ,\mbox{\bm $\pi$} \right)   \, ,
\label{eq:in1}
\end{eqnarray}
where $U(\sigma ,\mbox{\bm $\pi$})$ is the usual mexican-hat potential.
${\mathcal L}$  describes a system of interacting quarks, pions, sigmas and
a scalar-isoscalar chiral singlet field $\chi$.
The parameters of the model are: the chiral meson masses
$m_\pi=0.14$ GeV, $m_\sigma=1.2$ GeV, the pion decay constant
$f_\pi=93$ MeV, the quark--meson 
coupling constant $g$, and the mass $M$ of the $\chi$
field. 
The parameters $g$ and $M$, which are the only free parameters of
the model, are   fixed by 
reproducing the average nucleon-delta mass and the isoscalar radius
of the proton.
The technique used to compute the physical nucleon state 
$\vert N \rangle$ is based on a 
double projection of the mean-field solution
on linear and angular momentum eigenstates. 
It is a standard procedure and   
we refer the reader to \cite{NF93} for details about it. 

An important point to notice is that  
the intermediate states labeled by $\alpha$
in eq.~(\ref{deltaf}) are computed within the CDM in 
a parameter-free manner. The flavor asymmetries 
of the distribution functions arise from the differences between 
the intermediate states left out by a $u$ or a $d$ 
quark (antiquark) (the relevant formalism
can be found in Ref.\cite{thomas}) and therefore the 
results for these asymmetries 
are genuine predictions of the model. The two sources of the flavor  
asymmetry 
in this approach are therefore
the Pauli principle and the splitting of the masses of the intermediate
$4q$ states, due to pion exchange corrections. 
A crucial check of the reliability of our calculation 
comes from the fulfillment of the valence number sum rule, 
that we found to be saturated within few percent \cite{noi,h1}.
Another non-trivial test is provided by the 
Soffer inequality \cite{soffer}, 
which turns out to be satisfied by all quark 
and antiquark
distributions of our model.

Finally, we recall that 
the distributions computed in a quark model 
describe 
the nucleon at some low scale $\mu^2$ (the ``model scale'').   
They are used as the input
of the Altarelli--Parisi evolution from 
$\mu^2$ to a larger scale. 
In previous works \cite{noi,noi2} 
we  determined 
the model scale by comparing the model prediction for the 
valence momentum with the experimental value and found 
$\mu^2 = 0.16$ GeV$^2$.

\vspace{0.5cm}

\begin{figure}

\parbox{6cm}{
\scalebox{0.5}{
\includegraphics*[-30,30][700,690]{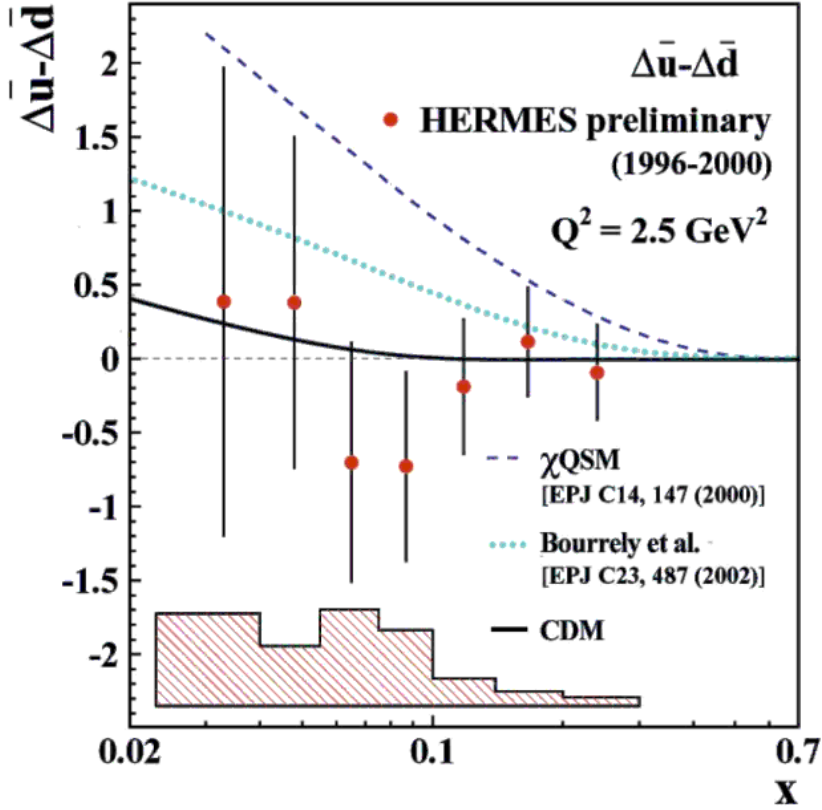}
}
}

\parbox{14cm}{
\caption{\label{difference} Flavor asymmetry 
of polarized sea in various models, compared with preliminary
data from HERMES. The error bars indicate statistical
errors, while the shaded band refers to systematic uncertainties.
Adapted from Fig. 4 of Ref.\cite{beckmann}.}}

\end{figure}

\begin{figure}

\parbox{6cm}{
\scalebox{0.7}{
\includegraphics*[30,430][530,730]{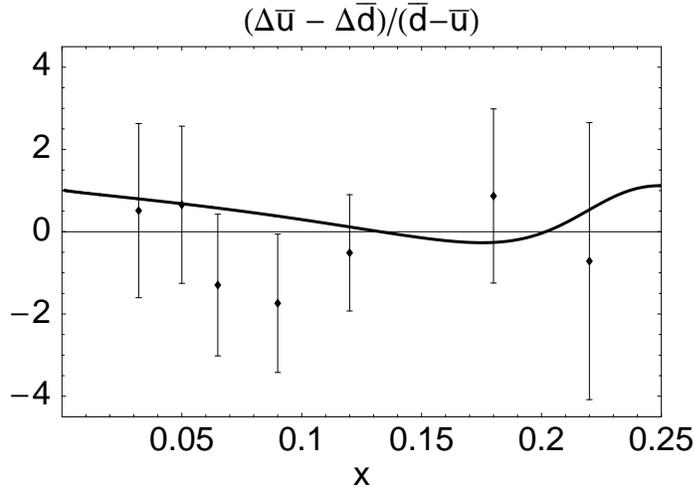}}
}

\parbox{14cm}{
\caption{\label{ratio1} Ratio of polarized 
to unpolarized isovector antiquark distributions computed
in the CDM model and compared with the ratio obtained from 
HERMES preliminary results
and the CTEQ5LO parametrization.}
}

\end{figure}

\noindent
3. We computed
various combinations of the isovector 
and isoscalar distributions (\ref{unpol}, \ref{pol}). 
Our result for the polarized flavor asymmetry 
$\Delta \bar f_- = \Delta \bar u - \Delta \bar d$, 
evolved in leading-order QCD
to the momentum scale of HERMES data, $Q^2 = 2.5$ GeV$^2$,
is shown in Fig.~\ref{difference} (solid line). 
We find that 
$\Delta \bar f_-$ 
is quite small and essentially zero for $x \gsim 0.1$. 
In Fig.~\ref{ratio1} we plot the ratio 
of the polarized to the unpolarized asymmetry,  
$(\Delta \bar u - \Delta \bar d)/
(\bar d - \bar u)$, at $Q^2 = 2.5$ GeV$^2$. This ratio is  
less than unity in the whole $x$ range (for 
the meaning of the data points in Fig.~\ref{ratio1}, see below).  
Finally, Fig.~\ref{ratio2} shows the isovector to isoscalar 
ratio  $\Delta \bar f_-/
\Delta  \bar f_+$ predicted by the CDM at the model scale $\mu^2$
and at the scale of the HERMES experiment.
This ratio must be taken with a grain of salt, 
since $\Delta \bar f_+$ 
has been evolved under the hypothesis of 
vanishing gluon polarization at the starting 
scale, which might be a simplistic assumption.

We must recall that
a negative feature of the CDM is that it 
yields single-quark wave functions 
that are very much peaked in momentum space.
Therefore quark 
distribution functions vanish too rapidly, 
typically above $x\sim 0.6$.
Also antiquark distributions vanish very fast.
However, we expect that 
the ratios presented in Figs.~2--3 should not be much affected 
by this behavior.

As mentioned earlier, 
the helicity densities have been also computed 
in the chiral quark soliton model (CQSM) \cite{diakonov,dressler,wakamatsu}. 
This model describes the nucleon as a state of 
$N_c$ valence quarks bound by a self-consistent 
hedgehog-like pion field. In the large-$N_c$ limit 
the distribution functions are 
calculated by a $1/N_c$ expansion. 
A clearcut  
prediction of the CQSM is that the isovector polarized 
antiquark distribution $\Delta \bar f_-$ is a leading 
quantity compared to the isoscalar distribution 
$\Delta \bar f_+$ and to the isovector 
unpolarized distribution $\bar f_-$, which both vanish 
at lowest order in $1/N_c$. Thus, one has in the CQSM
\be
\frac{\vert \Delta \bar f_- \vert}{\vert 
\Delta \bar f_+ \vert}
\sim N_c \gg 1\,, 
\;\;\;\;
\frac{\vert \Delta \bar f_- \vert}{\vert 
\bar f_- \vert}
\sim N_c \gg 1\,, 
\label{cqsm1}
\ee
and $\vert \Delta \bar f_- \vert$ is expected to be large. 
These behaviors are a direct consequence 
of the $1/N_c$ expansion and do not depend on the 
approximations used to calculate the distributions. 
The values of the two ratios 
(\ref{cqsm1}) can be read out from  
the results presented in \cite{wakamatsu} and \cite{dressler}, 
respectively. The order of 
magnitude is (the spread corresponds to 
the variation over the experimentally accessible $x$ range)
\be
\frac{\Delta \bar u - \Delta \bar d}{\Delta \bar 
u + \Delta \bar d} \sim 2-3\,, 
\;\;\;\;
\frac{\Delta \bar u - \Delta \bar d}{\bar 
d - \bar u} \sim 3-4\,
\label{cqsm2}
\ee
The differences between the CDM and the CQSM   
predictions are therefore very large 
and can be fully appreciated 
in Fig.~1. 
The observables plotted 
in the three figures are extremely sensitive 
to the model used for computing them, and   
their accurate experimental determination 
would allow a definite test of the theory.

\begin{figure}[t]

\parbox{6cm}{
\scalebox{0.7}{
\includegraphics*[90,400][580,740]{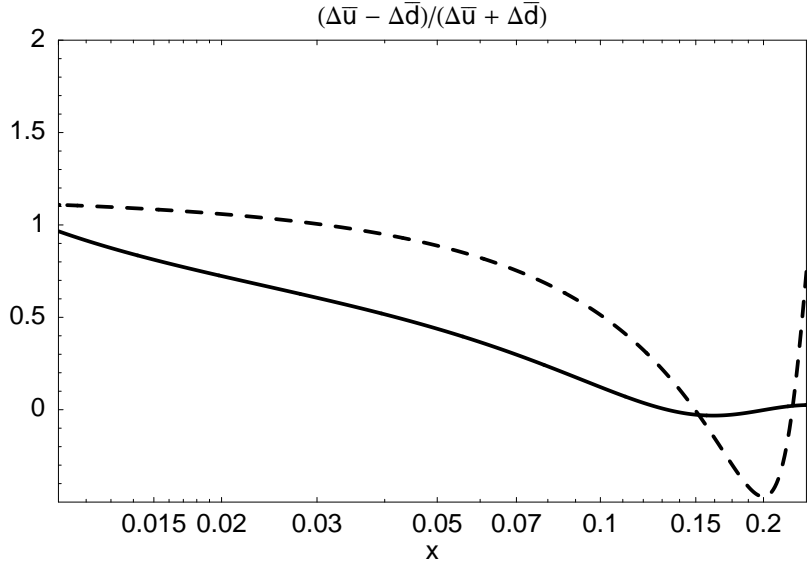}}
}

\parbox{14cm}{
\caption{\label{ratio2} Ratio of 
isovector to isoscalar polarized sea distributions computed
in the CDM model, at the scale of the model (dashed line) and evolved to
$Q^2=2.5$ GeV$^2$ (solid line).}
}

\end{figure}

Let us take a look at the available data.
The quantity measured by HERMES is the 
semi-inclusive cross section asymmetry $A_1^h$. 
In leading-order QCD and under the assumption 
that the transverse spin structure function 
$g_2$ vanishes, $A_1^h$ reads
\be
A_1^h(x, Q^2) = \frac{1 + R(x, Q^2)}{1 + \gamma^2} 
\, \frac{\sum_f e_f^2 \Delta f(x, Q^2) \int 
D_f^h(z, Q^2) \, {\rm d} z}{\sum_f e_f^2 f(x, Q^2) \int 
D_f^h(z, Q^2) \, {\rm d} z}\,. 
\label{a1h}
\ee
Here $R = \sigma_L/\sigma_T$ is the longitudinal to transverse 
photo-absorption cross section ratio, $\gamma = 2 m_N x/Q$, 
and $D_f^h$ are the fragmentation functions of flavor 
$f$ into an hadron $h$ carrying a fraction $z$ of the initial quark momentum. 
By combining data on hydrogen and deuterium 
targets and detecting final state pions and kaons in the range
$0.2<z<0.8$,    
HERMES extracted separately $\Delta u$, $\Delta \bar u$, 
$\Delta d$, $\Delta \bar d$ and $\Delta s + \Delta \bar s$. 
As shown in Fig.~\ref{difference}, the 
isovector antiquark distribution $\Delta \bar f_- 
= \Delta \bar u - \Delta \bar d$ 
is found to be small and compatible with 
zero. This is still a preliminary 
result, affected by significant statistical and systematic errors, but it is   
reproduced reasonably well by our model 
(see Fig.~\ref{difference}). On the contrary, 
the large value of $\Delta \bar f_-$ predicted by 
the CQSM (dashed curve in Fig.~\ref{difference}) seems 
to be discarded by the data (the dotted curve 
is the prediction of the statistical model 
of Bourrely {\it et al.} \cite{bourrely}, that 
we do not discuss here). In Fig.~\ref{ratio1} 
we divided the HERMES data by the 
unpolarized asymmetry $\bar d - \bar u$ 
as given by the CTEQ5LO parameterization \cite{cteq},
which is essentially driven by the Drell-Yan data.  
We attributed to the CTEQ5LO fit an 
absolute uncertainty of $\sim 10\%$
in the relevant range of $x$, mimicking in this way the Drell-Yan errors. 
The resulting points are plotted with the propagated 
errors and the total error bars are dominated by
the large uncertainties on  
$\Delta \bar u - \Delta \bar d$. Once more, the agreement with 
the CDM prediction (solid line) is fairly good, 
while the high CQSM values seem to be excluded.

\vspace{0.5cm}

\noindent
4. Before coming to the conclusions, we would 
like to comment on some 
technicalities concerning the model calculation 
of antiquark densities. Let us first notice
that, if we adhere to the definition (\ref{defhel1}) of quark 
distributions, the variable $x= k^+/p^+$ (where $k^+$ 
is the light-cone quark momentum) is not constrained 
a priori to be positive. It turns out that there is a relation connecting 
quark and antiquark distributions, which are obtained 
by continuing $x$ to negative values.  
For helicity distributions this relation is  
\be
\Delta \bar f (x) = \Delta f(-x)\,. 
\label{relqqbar}
\ee
In some approaches, including that of \cite{diakonov,dressler,wakamatsu}, 
the antiquark distributions are computed by 
means of (\ref{relqqbar}). This is, in principle,  
an unsafe procedure. The reason is that 
there are semi-connected diagrams that contribute 
to the distributions for $x <0$, whereas in computing 
these distributions in the physical region only 
connected diagrams should be considered (indeed, this {\em defines} 
the parton model, as pointed out by Jaffe \cite{jaffe}).
Our approach has no such problem: the antiquark 
distributions are calculated directly from their 
field-theoretical expression (\ref{defhel2}), 
by inserting a complete   
set of intermediate states, as explained above. 
Incidentally, we notice that the different techniques adopted for computing
the antiquark distributions are probably at the origin of the 
sign discrepancy between the transversity sea distributions 
computed in \cite{h1} and in \cite{wakamatsu}.

\vspace{0.5cm}

\noindent
5. In summary, we showed that the chiral 
chromodielectric model and the 
chiral quark soliton model predict very different 
behaviors for the polarized isovector  
distribution $\Delta \bar f_-$ and for the ratios 
of $\Delta \bar f_-$ to the unpolarized 
isovector distribution, $\Delta \bar f_-/\bar f_-$, 
and to the polarized isoscalar distribution, 
$\Delta \bar f_-/\Delta \bar f_+$. The
recent preliminary HERMES data favor 
the CDM results and exclude the large   
value for $\Delta \bar f_-$ predicted by the CQSM.  
Hopefully, more precise data in the next 
future will say a conclusive 
word about the whole question.

\vspace{1cm}
It is a pleasure to thank Paola Ferretti Dalpiaz for various 
useful discussions.


\end{document}